\documentclass[twocolumn,superscriptaddress,showpacs,prl]{revtex4}

\usepackage[dvips]{graphicx,color}
\usepackage{delarray}
\usepackage{ogonek}
\usepackage{amsmath}
\usepackage{bm}

\newcommand{\ket}[1]{| { #1} \rangle}
\newcommand{\bra}[1]{ \langle {#1} |}

\begin{document}
\title{Aggregating quantum repeaters for the quantum internet}

\author{Koji Azuma}
\email{azuma.koji@lab.ntt.co.jp}
\affiliation{NTT Basic Research Laboratories, NTT Corporation, 3-1 Morinosato Wakamiya, Atsugi, Kanagawa 243-0198, Japan}

\author{Go Kato}
\email{kato.go@lab.ntt.co.jp}
\affiliation{NTT Communication Science Laboratories, NTT Corporation, 3-1 Morinosato Wakamiya, Atsugi, Kanagawa 243-0198, Japan}

\
\date{\today}

\begin{abstract}
The quantum internet holds promise for performing quantum communication, such as quantum teleportation and quantum key distribution, freely between any parties all over the globe. 
For such a quantum internet protocol, a general fundamental upper bound on the performance has been derived [K.~Azuma, A.~Mizutani, and H.-K.~Lo, arXiv:1601.02933]. Here we consider its converse problem.
In particular, we present a protocol constructible from any given quantum network, which is based on running quantum repeater schemes in parallel over the network.
The performance of this protocol and the upper bound restrict the quantum capacity and the private capacity over the network from both sides. 
The optimality of the protocol is related to fundamental problems such as additivity questions for quantum channels and questions on the existence of a gap between quantum and private capacities.
\pacs{03.67.Hk, 03.67.Dd, 03.65.Ud, 03.67.-a}
\end{abstract}
\maketitle


In the Internet, if a client communicates with a far distant client, 
the data travel across multiple networks.
At present, the nodes and the communication channels in the networks are composed of physical devices governed by the laws of classical information theory,
and the data flow obeys the celebrated max-flow min-cut theorem in graph theory.
However, in the future, such classical nodes and channels should be replaced with quantum ones, whose network follows the rules of quantum information theory, rather than classical one.
This network, called {\it quantum internet}, could accomplish tasks that are intractable in the realm of classical information processing,
and it serves opportunities and challenges across a range of intellectual and technical frontiers, including quantum communication, computation, metrology, and simulation \cite{K08}. So far, the main interest in the quantum internet has been its realization \cite{B98,DLCZ,SSRG09,ATKI10,M10,M12,J09,C06,L06,G12,L12,ZDB12,KWD03,MATN15,ATL15}. 
But, it must be one of the most fundamental trials from a theoretical perspective to grasp the full potential of the quantum internet. Along this line, recently, a general fundamental upper bound on the performance was derived \cite{AML16} for its use for supplying two clients with entanglement or a secret key.
Interestingly, this upper bound is estimable and applied to any private-key or entanglement distillation scheme that works over any network topology composed of arbitrary quantum channels by using arbitrary local operations and unlimited classical communication (LOCC). 
With this, for the case of linear lossy optical channel networks, it has been shown \cite{AML16} that existing intercity quantum key distribution (QKD) protocols \cite{ATM15,AKB14,PRML14} and quantum repeater schemes \cite{M12,ATL15,M10,J09,G12} have no scaling gap with the fundamental upper bound.
Moreover, in the case of a multipath network composed of a wide range of stretchable quantum channels (including lossy optical channels), 
it has been proven \cite{P16} to be optimal to choose a {\it single} path between two clients for running quantum repeater scheme, in order to
minimize the number of times paths between them are used to obtain a secret key or entanglement.

In this paper, we consider a general converse problem inspired by the form \cite{AML16} of the fundamental upper bound. 
In particular, we provide a protocol constructible from any given quantum network, which runs quantum repeater schemes in parallel over the network to provide entanglement or a secret key to two clients.
The performance of this protocol and the upper bound are in the same form as represented by the left-hand and right-hand sides of Eqs.~(\ref{eq:asym-1}) and (\ref{eq:asym-2}), restricting the quantum capacity and the private capacity over the network from both sides. 
Especially, in the case of the lossy optical channel network, 
our protocol is shown to have no scaling gap with the upper bound, irrespectively of the network topology.
The optimality of the protocol is indeed related to fundamental problems such as additivity questions for quantum channels and questions on the existence of a gap between quantum and private capacities.
Since these problems were solved \cite{PLO15} by Pirandola {\it et al.} for an important class of practical quantum channels with stretchability, our protocol is optimal for networks composed of such practical channels.

\begin{figure}[b] 
\includegraphics[keepaspectratio=true,height=33mm]{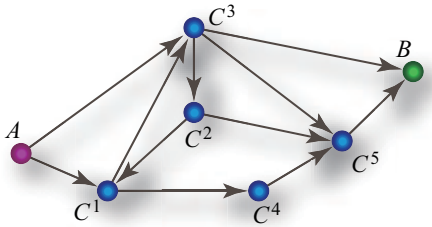}
\caption{Quantum network. The network is associated with a directed graph $G=(V,E)$ with set $V$ of vertices and set $E$ of edges, where $V$ is composed of Alice's node $A$, Bob's node $B$ and intermediate nodes $C^1,C^2,\ldots$, and $C^n$ ($n=5$ here) and a directed edge $e=X \to Y$ in $E$ for $X, Y \in V$ specifies a quantum channel ${\cal N}^e$ to send a subsystem in node $X$ to node $Y$. The goal here is to give Alice and Bob resources for quantum communication, secret bits or ebits, by using quantum channels $\{{\cal N}^e\}_{e\in E}$ and LOCC.}
\label{fig:1}
\end{figure}

{\it Quantum internet protocol for two clients.}---Let us begin by introducing quantum internet protocols for two clients and by reviewing the fundamental upper bound for them \cite{AML16}. A quantum internet protocol will serve a subnetwork to two clients, called Alice and Bob, to provide resources for quantum communication, secret bits or ebits. The subnetwork is associated with a directed graph $G=(V,E)$ with set $V$ of vertices and set $E$ of edges (see Fig.~\ref{fig:1} as an example), where $V$ is composed of Alice's node $A$, Bob's node $B$ and intermediate nodes $C^1,C^2,\ldots$, and $C^n$ and an edge $e=X \to Y$ in $E$ for $X, Y \in V$ specifies a quantum channel ${\cal N}^e$ to send a subsystem in node $X$ to node $Y$.
In general, the protocol begins by sharing a separable state and then by using a quantum channel ${\cal N}^{e_1}$ with $e_1\in E$. This is followed by LOCC among all the nodes, giving an outcome $k_1$ and a quantum state $\hat{\rho}^{ABC^1C^2\ldots C^n}_{k_1}$ with probability $p_{k_1}$. In the $i$-th round, depending on the previous outcome ${\bm k}_{i-1}=k_{i-1}\ldots k_2 k_1$ (with ${\bm k}_0:=1$), the protocol uses a quantum channel ${\cal N}^{e_{{\bm k}_{i-1}}}$ with $e_{{\bm k}_{i-1}}\in E$, followed by LOCC providing a quantum state $\hat{\rho}_{{\bm k}_i}^{ABC^1C^2\ldots C^n}$ corresponding to an outcome $k_i$ with probability $p_{k_i|{\bm k}_{i-1}}$.
In a final round, say an $l$-th round, the protocol provides a quantum state $\hat{\rho}_{{\bm k}_l}^{AB}:={\rm Tr}_{C^1C^2\ldots C^n}(\hat{\rho}_{{\bm k}_l}^{ABC^1C^2\ldots C^n})$ close to a target state $\hat{\tau}^{AB}_{d_{{\bm k}_l}}$ in the sense $\|\hat{\rho}_{{\bm k}_l}^{AB}-\hat{\tau}^{AB}_{d_{{\bm k}_l}}\|_1\le \epsilon$ with $\epsilon >0$, from which 
$\log_2 d_{{\bm k}_l}$ ebits for quantum teleportation or $\log_2 d_{{\bm k}_l}$ secret bits for unconditionally secure communication are distilled.

\begin{figure}[b] 
\includegraphics[keepaspectratio=true,height=33mm]{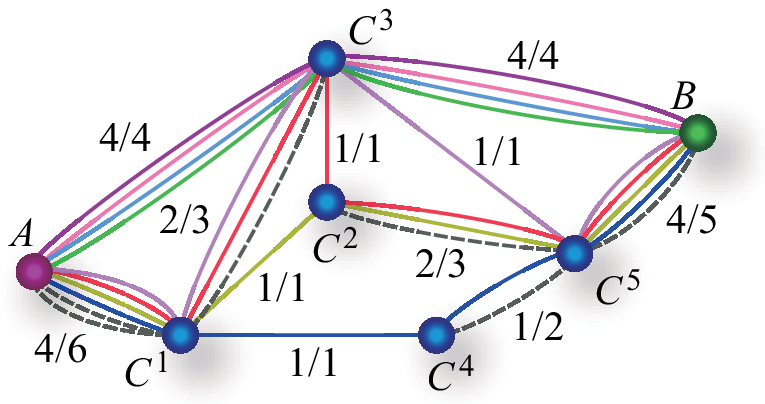}
\caption{Bell-pair network $\bigotimes_{e\in E} \ket{\Phi^+}\bra{\Phi^+}^{\otimes \lfloor \bar{l}^e\rfloor R^e_\varepsilon } _e$ associated with $G'=(V,E')$. This is an example of the Bell-pair network for the quantum network in Fig.~\ref{fig:1}. Each undirected edge $e'\in E'$ represents a Bell pair $\ket{\Phi^+}_{e}$ generated by a maximal-entanglement distribution protocol. The denominator and the numerator of a fraction describe $ \lfloor \bar{l}^e\rfloor R^e_\varepsilon$ and how many Bell pairs are utilized in the aggregated quantum repeater protocol, respectively. Dashed edges are unused Bell pairs. Here, the choice of $V_A=\{A,C^1,C^3\}$ in Eq.~(\ref{eq:M}) gives $M_\varepsilon=8$.}
\label{fig:2} 
\end{figure}

{\it Single-letter general upper bound.}---For the quantum internet protocol, a general upper bound on the performance has been given \cite{AML16}, which is described as follows.
Let us divide set $V$ into two disjoint sets, $V_A$ including $A$ and $V_B$ including $B$, satisfying $V_B=V \setminus V_A$.
If ${\cal N}^{e_{{\bm k}_i}}$ is a channel between a node in $V_A$ and a node in $V_B$, we write ${\bm k}_i \in K_{V_A\leftrightarrow V_B}$.
Then, the most general protocol has a limitation described by
\begin{multline}
\sum_{{\bm k_l} } p_{{\bm k}_l} \log_2 d_{{\bm k}_l} 
\le \frac{1}{1-16\sqrt{\epsilon}} \\ \times \left(\min_{V_A} \sum_{i=0}^{l-1} \sum_{{\bm k_{i}} \in K_{V_A\leftrightarrow V_B} } p_{{\bm k}_{i}} E_{\rm sq} ( {\cal N}^{e_{{\bm k}_{i}}}) + 
4 h(2\sqrt{\epsilon}) \right), \label{eq:mc}
\end{multline}
where $p_{{\bm k}_i}:=p_{k_i| {\bm k}_{i-1}} \ldots p_{k_3| {\bm k}_2} p_{k_2|{\bm k}_1} p_{k_1}$,
$h$ is the binary entropy function with a property of $\lim_{x \to 0} h(x)=0$. 
$E_{\rm sq}({\cal N}^{X\to Y})$ is the squashed entanglement of the channel \cite{TGW14,TGW14e} defined by
\begin{equation}
E_{\rm sq}({\cal N}^{X\to Y})=\max_{\ket{\phi}_{xx'}} E_{\rm sq}^{x:y} ({\cal N}^{X\to Y}(\ket{\phi}\bra{\phi}_{xx'}))
\end{equation}
for channel ${\cal N}^{X\to Y}$ to output a system $y$ for a party $Y$ for the input subsystem $x'$ of a party $X$,
where $E_{\rm sq}^{x:y} $ is the squashed entanglement \cite{CW04} between system $x$ of party $X$ and system $y$ of party $Y$.
This implies that the bound (\ref{eq:mc}) is a single-letter formula, that is, it can be evaluated as a function of a single channel use.
We also note that the bound (\ref{eq:mc}) is reduced to $\sum_{{\bm k_l} } p_{{\bm k}_l} \log_2 d_{{\bm k}_l} \le \min_{V_A}\sum_{i=0}^{l-1} \sum_{{\bm k_{i}} \in K_{V_A\leftrightarrow V_B} } p_{{\bm k}_{i}} E_{\rm sq} ( {\cal N}^{e_{{\bm k}_{i}}}) $ for $\epsilon \to 0$.

{\it Converse problem.}---The bound (\ref{eq:mc}) suggests that a given protocol can also be characterized as follows.
Let $\langle f_{{\bm k}_i} \rangle_{{\bm k}_i}$ be the average of function $f_{{\bm k}_i}$ over ${\bm k}_i$, i.e., $\langle f_{{\bm k}_i} \rangle_{{\bm k}_i}=\sum_{{\bm k}_i} p_{{\bm k}_i} f_{{\bm k}_i}$.
For $e\in E$,
$\bar{l}^e:=\sum_{i=0}^{l-1} \langle \delta_{e,e_{{\bm k}_i}} \rangle_{{\bm k}_i}$ with the Kronecker delta $\delta$ represents the average number of times quantum channel ${\cal N}^{e}$ is used. If $e \in V_A \leftrightarrow V_B$ denotes that the edge $e$ has the tail belonging to $V_A$ ($V_B$) and the head belonging to $V_B$ ($V_A$),
the bound (\ref{eq:mc}) can be rephrased as 
\begin{multline}
\langle \log_2 d_{{\bm k}_l} \rangle_{{\bm k}_l} 
\le \frac{1}{1-16\sqrt{\epsilon}} \\ \times \left(\min_{V_A} \sum_{e \in V_A\leftrightarrow V_B} 
\bar{l}^e E_{\rm sq}({\cal N}^e)+ 
4 h(2\sqrt{\epsilon}) \right). \label{eq:mc-m}
\end{multline}
Hence, the bound depends only on the set $\{\bar{l}^e,E_{\rm sq}({\cal N}^e)\}_{e \in E}$ determined by the given protocol, implying that the protocol is generally characterized by how many times given quantum channels $\{{\cal N}^e\}_{e \in E}$ are used in the protocol.
Therefore, the converse problem is to find, if any, a protocol that saturates relation (\ref{eq:mc-m}) by using 
quantum channel ${\cal N}^{e}$, at most, $\bar{l}^e$ times on average for $e\in E$.
The main point of our paper is to present a candidate of such a protocol.

{\it Aggregated quantum repeater protocol.}---We introduce a protocol, referred to as an {\it aggregated quantum repeater protocol}, that runs quantum repeater protocols in parallel over the quantum network by using quantum channels $\{ ({\cal N}^{e})^{\otimes \lfloor \bar{l}^e \rfloor } \}_{e\in E}$, where $\lfloor z \rfloor$ represents the largest integer $\le z$.
Our protocol begins by running a maximal-entanglement distribution protocol between nodes connected by quantum channel ${\cal N}^{e}$. 
In particular, the protocol starts by sending a half of a bipartite system through the quantum channel $({\cal N}^e)^{\otimes \lfloor \bar{l}_e\rfloor}$, and then performs an entanglement distillation protocol between the two ends of edge $e$. 
Suppose that this protocol provides a state $\hat{\rho}^e$ close to $ \lfloor \bar{l}^e\rfloor R^e_{\varepsilon} $ copies of a Bell pair $\ket{\Phi^+}_e$, i.e.,
$\|\hat{\rho}^e-\ket{\Phi^+}\bra{\Phi^+}^{\otimes \lfloor \bar{l}^e\rfloor R_\varepsilon^e }_e\|_1\le \varepsilon$ with $\varepsilon>0$.
By running this protocol all over the edges $e\in E$, we obtain a state $\bigotimes_{e\in E} \hat{\rho}^e$ with 
\begin{equation}
\Big\|\bigotimes_{e\in E} \hat{\rho}^e -\bigotimes_{e\in E} \ket{\Phi^+}\bra{\Phi^+}^{\otimes \lfloor \bar{l}^e \rfloor R^e_\varepsilon}_e \Big\|_1 \le |E| \varepsilon, \label{eq:error}
\end{equation}
where $|E|$ is the cardinality of set $E$.
Let us regard each of the Bell pairs $\bigotimes_{e\in E} \ket{\Phi^+}\bra{\Phi^+}^{\otimes \lfloor \bar{l}^e\rfloor R^e_\varepsilon }_e$, a Bell pair $\ket{\Phi^+}_e$ for instance, as an undirected edge $e'$ with two ends of $e$,
and let $E'$ be the set composed of all such edges $e'$.
Then, the Bell-pair network, i.e., $\bigotimes_{e\in E} \ket{\Phi^+}\bra{\Phi^+}^{\otimes \lfloor \bar{l}^e\rfloor R^e_\varepsilon }_e$, can be associated with an undirected graph defined by $G':=(V,E')$ (see Fig.~\ref{fig:2} as an example).
Here we invoke Menger's theorem in graph theory.

{\it Menger's theorem (Edge version)} \cite{BM}.---In any graph $G$ with two distinguished vertices $A$ and $B$, the maximum number of pairwise edge-disjoint $AB$-paths is equal to the minimum number of edges in an $AB$-cut.

Let $M_\varepsilon$ be the minimum number of edges in an $AB$-cut in the graph $G'$, i.e.,
\begin{equation}
M_\varepsilon:=\min_{V_A} \sum_{e \in V_A \leftrightarrow V_B} \lfloor \bar{l}^e\rfloor R^e_\varepsilon . \label{eq:M}
\end{equation}
Then, 
Menger's theorem states that there are $M_\varepsilon$ pairwise edge-disjoint $AB$-paths in graph $G'$ (see Fig.~\ref{fig:2} for example). Since each $P_i$ of these $AB$-paths $\{P_i\}_{i=1,2,\ldots,M_\varepsilon}$ corresponds to a linear chain of Bell pairs in the Bell-pair network $\bigotimes_{e\in E} \ket{\Phi^+}\bra{\Phi^+}^{\otimes \lfloor \bar{l}^e\rfloor R^e_\varepsilon}_e$, the linear chain can be transformed into a Bell pair $\ket{\Phi^+}_{AB}$ by performing entanglement swapping ${\cal S}_{P_i}$ (including a Pauli correction) over the intermediate nodes on $P_i$.
Then, from Eq.~(\ref{eq:error}), 
we have
\begin{align}
|E| \varepsilon & \ge
\Big\|\bigotimes_{e\in E} \hat{\rho}^e -\bigotimes_{e\in E} \ket{\Phi^+}\bra{\Phi^+}^{\otimes \lfloor \bar{l}^e\rfloor R^e_\varepsilon }_e \Big\|_1 \nonumber \\
& \ge \Big\| {\rm Tr}_{C^1C^2\ldots C^n} \circ {\cal S} \Big( \bigotimes_{e\in E} \hat{\rho}^e-\bigotimes_{e\in E} \ket{\Phi^+}\bra{\Phi^+}^{\otimes \lfloor \bar{l}^e\rfloor R^e_\varepsilon }_e\Big) \Big\|_1 \nonumber \\
& = \Big\| \hat{\rho}^{AB} - \ket{\Phi^+}\bra{\Phi^+}_{AB}^{\otimes M_\varepsilon} \Big\|_1 ,
\end{align}
where ${\cal S}:={\cal S}_{P_{M_\varepsilon}} \circ \ldots \circ {\cal S}_{P_2} \circ {\cal S}_{P_1}$ and $ \hat{\rho}^{AB} :={\rm Tr}_{C^1C^2\ldots C^n} \circ {\cal S} ( \bigotimes_{e\in E} \hat{\rho}^e)$. Therefore, the protocol, just like aggregating quantum repeater protocols, provides $M_\varepsilon$ ebits or secret bits by using quantum channels $\{ ({\cal N}^{e})^{\otimes \lfloor \bar{l}^e \rfloor } \}_{e\in E}$ with error $|E| \varepsilon $.

{\it Bounds on the optimal $\epsilon$-close protocol.}---To evaluate the performance of the aggregated quantum repeater protocol, we begin by introducing the concept of $\epsilon$-close protocols.
If a protocol presents $\langle \log_2 d_{{\bm k}_l} \rangle_{{\bm k}_l} $ ebits or secret bits with an error $\le \epsilon$ (in terms of the trace distance) by using 
quantum channel ${\cal N}^{e}$, at most, $\bar{l}^e$ times on average for $e\in E$, we call the protocol an $\epsilon$-close protocol.
Let ${\cal P}_{\epsilon}$ be the set of all the $\epsilon$-close protocols.
Then, the aggregated quantum repeater protocol with error $|E| \varepsilon \le \epsilon$
and the bound (\ref{eq:mc-m}) show 
\begin{multline}
\min_{V_A} \sum_{e \in V_A \leftrightarrow V_B} \lfloor \bar{l}^e \rfloor R_\varepsilon^e \le \sup_{{\cal P}_\epsilon} \langle \log_2 d_{{\bm k}_l} \rangle_{{\bm k}_l}\\
\le
\frac{1}{1-16\sqrt{\epsilon}} \left(\min_{V_A} \sum_{e \in V_A\leftrightarrow V_B} 
\bar{l}^e E_{\rm sq}({\cal N}^e)+ 
4 h(2\sqrt{\epsilon})\right). \label{eq:main}
\end{multline}
This shows that the best performance of the $\epsilon$-close protocols is sandwiched between the performance of the aggregated quantum repeater protocol and the fundamental upper bound.

{\it Asymptotic limits.}---Let us consider the asymptotic limits of (\ref{eq:main}).
We first introduce frequency $\bar {f}^e:=\bar {l}^e/l$. 
For $\bar{f}^e>0$, 
$\bar{l}^e \to \infty$ for the limit $l\to \infty$, for which, by optimizing the maximal-entanglement distribution protocol, $R^e_\varepsilon$ can reach the quantum capacity $ Q^\leftrightarrow({\cal N}^e) $ of channel ${\cal N}^e$ assisted by unlimited forward and backward classical communication in the limit $\epsilon\to 0$.
Therefore, in this asymptotic limit, the inequalities (\ref{eq:main}) are reduced to 
\begin{multline}
\min_{V_A} \sum_{e \in V_A \leftrightarrow V_B} \bar{f}^e Q^\leftrightarrow({\cal N}^e) 
\le \lim_{\epsilon \to 0} \lim_{l\to \infty} \sup_{{\cal P}_\epsilon} \frac{ 
\langle \log_2 d_{{\bm k}_l} \rangle_{{\bm k}_l}}{l}\\
\le \min_{V_A} \sum_{e \in V_A\leftrightarrow V_B} 
\bar{f}^e E_{\rm sq}({\cal N}^e). \label{eq:asym-1}
\end{multline}
Hence, the minimum $AB$-cuts over functions $Q^\leftrightarrow$ and $E_{\rm sq}$ defined on the quantum channel network $\{{\cal N}^e\}_{e \in E}$ restrict
the quantum communication capacity and the private capacity per total channel use from both sides.

We can also consider another asymptotic limit on the capacities per time.
Suppose that we use quantum channel ${\cal N}^e$, at most, $l^e$ times for time $t$.
Let us introduce rate $\bar{r}^e:=\bar{l}^e/t$. Then, with $l=t\sum_e \bar{r}^e$, we have another asymptotic limit of Eq.~(\ref{eq:main}) as
\begin{multline}
\min_{V_A} \sum_{e \in V_A \leftrightarrow V_B} \bar{r}^e Q^\leftrightarrow({\cal N}^e) 
\le \lim_{\epsilon \to 0} \lim_{t\to \infty} \sup_{{\cal P}_\epsilon} \frac{ 
\langle \log_2 d_{{\bm k}_l} \rangle_{{\bm k}_l}}{t}\\
\le \min_{V_A} \sum_{e \in V_A\leftrightarrow V_B} 
\bar{r}^e E_{\rm sq}({\cal N}^e). \label{eq:asym-2}
\end{multline}

The right-hand sides of Eqs.~(\ref{eq:asym-1}) and (\ref{eq:asym-2}) are the bounds applied to arbitrary protocols that may include even multi-party entanglement purification and quantum network coding. However, these bounds are written by the single-letter quantity, i.e., the squashed entanglement of the channel.
On the other hand, the left-hand sides of Eqs.~(\ref{eq:asym-1}) and (\ref{eq:asym-2}) are the bounds for the aggregated quantum repeater protocol that does not use such multi-party protocols at all. Besides, the bounds are merely described by using the quantum capacity that is intractable to be estimated in general.
Despite these differences, the symmetric form of left-hand and right-hand sides of Eqs.~(\ref{eq:asym-1}) and (\ref{eq:asym-2}) may suggest that the aggregated quantum repeater protocol could work very efficiently.

{\it Capacities of lossy optical channel networks.}---Let us consider a quantum network composed of lossy optical channels as an example of quantum networks, in order to see how efficiently the aggregated quantum repeater protocol could work.
For a lossy optical channel ${\cal N}_{\eta}$ with transmittance $\eta$, its quantum capacity $ Q^\leftrightarrow({\cal N}_\eta) $ \cite{PLO15} and an upper bound \cite{TGW14} of the squashed entanglement $E_{\rm sq}({\cal N}_\eta) $ per mode have been derived as follows:
\begin{align} 
Q^\leftrightarrow({\cal N}_\eta) &= \log_2\left(\frac{1}{1-\eta} \right) , \label{eq:lossy-1}\\
E_{\rm sq} ({\cal N}_\eta) &\le \log_2\left(\frac{1+\eta}{1-\eta} \right) . \label{eq:lossy-2}
\end{align}
Hence, if we consider a network with ${\cal N}^e = {\cal N}_{\eta^e}$ for $e\in E$, from Eqs.~(\ref{eq:lossy-1}) and (\ref{eq:lossy-2}), the bound (\ref{eq:asym-1}) becomes
\begin{multline}
\min_{V_A} \sum_{e \in V_A \leftrightarrow V_B} \bar{f}^e \log_2\left(\frac{1}{1-\eta^e} \right) \\
\le \lim_{\epsilon \to 0} \lim_{l\to \infty} \sup_{{\cal P}_\epsilon} \frac{ 
\langle \log_2 d_{{\bm k}_l} \rangle_{{\bm k}_l}}{l}\\
\le \min_{V_A} \sum_{e \in V_A\leftrightarrow V_B} 
\bar{f}^e \log_2\left(\frac{1+\eta^e}{1-\eta^e} \right) . \label{eq:lossy}
\end{multline}

Inequalities (\ref{eq:lossy}) show that the aggregated quantum repeater protocol has no scaling gap with a best protocol. In fact,
since $\max_{e\in V_A\leftrightarrow V_B} E_{\rm sq} ({\cal N}_{\eta^e}) /Q^\leftrightarrow({\cal N}_{\eta^e}) \le \max_{e\in E} E_{\rm sq} ({\cal N}_{\eta^e}) /Q^\leftrightarrow({\cal N}_{\eta^e}) \le 2$, we have 
\begin{equation}
\sum_{e \in V_A\leftrightarrow V_B} 
\bar{f}^e \log_2\left(\frac{1+\eta^e}{1-\eta^e} \right) \le 2 \sum_{e \in V_A \leftrightarrow V_B} \bar{f}^e \log_2\left(\frac{1}{1-\eta^e} \right) .
\end{equation}
Hence, from Eq.~(\ref{eq:lossy}), we obtain
\begin{multline}
\min_{V_A} \sum_{e \in V_A \leftrightarrow V_B} \bar{f}^e \log_2\left(\frac{1}{1-\eta^e} \right) \\
\le \lim_{\epsilon \to 0} \lim_{l\to \infty} \sup_{{\cal P}_\epsilon} \frac{ 
\langle \log_2 d_{{\bm k}_l} \rangle_{{\bm k}_l}}{l}\\
\le 2 \min_{V_A} \sum_{e \in V_A \leftrightarrow V_B} \bar{f}^e \log_2\left(\frac{1}{1-\eta^e} \right) .\label{eq:b-lossy}
\end{multline}
Since the left-hand side of these inequalities represents the performance of the aggregated quantum repeater protocol, this shows that 
the aggregated quantum repeater protocol has no scaling gap with a best protocol irrespectively of the optical channel network topology. 
More importantly, Eq.~(\ref{eq:b-lossy}) proves that, although a best protocol may use multi-party entanglement purification or quantum network coding, the communication performance is merely, at most, twice of that of the aggregated quantum repeater protocol based only on 
linear networks. 

{\it On the optimality of the aggregated quantum repeater protocol.}---By considering the converse problem that has been induced by the general bound (\ref{eq:mc}),
we have found an aggregated quantum repeater protocol that is based on running quantum repeater protocols in parallel over the quantum network. In the case of the lossy optical channel network, the repeater protocol has been shown to have no scaling gap with the general bound (\ref{eq:mc}) as well as the best protocol.
However, if we consider an alternative upper bound---inspired by the conception in the derivation \cite{AML16} of the bound (\ref{eq:mc})---for the case of the asymptotic limit, the aggregated quantum repeater protocol could indeed be optimal. To show this, let $Q^\leftrightarrow (\{\bar{f}^e/\bar{f}^{V_A \leftrightarrow V_B}, {\cal N}^{e} \}_{e\in V_A \leftrightarrow V_B})$ ($K^\leftrightarrow (\{\bar{f}^e/\bar{f}^{V_A \leftrightarrow V_B}, {\cal N}^{e} \}_{e\in V_A \leftrightarrow V_B})$) with $\bar{f}^{V_A \leftrightarrow V_B}:=\sum_{e\in V_A \leftrightarrow V_B} \bar{f}^e $ denote the quantum capacity (private capacity) defined under the limit of $l \to \infty$ for the paradigm where parties $V_A$ and $V_B$ are allowed to use quantum channels ${\cal N}^e$ between them (i.e., $e \in V_A \leftrightarrow V_B$) $\lfloor l \bar{f}^e \rfloor$ times and their LOCC in order to distill ebits (secret bits).
Since any quantum internet protocol can be regarded as a bipartite protocol between $V_A$ and $V_B$ \cite{AML16}, we have
\begin{multline}
\lim_{\epsilon \to 0} \lim_{l\to \infty} \sup_{{\cal P}_\epsilon} \frac{ 
\langle \log_2 d_{{\bm k}_l} \rangle_{{\bm k}_l}}{l} \\
\le \min_{V_A} \bar{f}^{V_A \leftrightarrow V_B} Q^\leftrightarrow (\{\bar{f}^e/\bar{f}^{V_A \leftrightarrow V_B}, {\cal N}^{e} \}_{e\in V_A \leftrightarrow V_B}) \label{eq:upp-1}
\end{multline}
for entanglement distillation for Alice and Bob and 
\begin{multline}
\lim_{\epsilon \to 0} \lim_{l\to \infty} \sup_{{\cal P}_\epsilon} \frac{ 
\langle \log_2 d_{{\bm k}_l} \rangle_{{\bm k}_l}}{l} \\
\le \min_{V_A} \bar{f}^{V_A \leftrightarrow V_B} K^\leftrightarrow (\{\bar{f}^e/\bar{f}^{V_A \leftrightarrow V_B}, {\cal N}^{e} \}_{e\in V_A \leftrightarrow V_B}) \label{eq:upp-2}
\end{multline}
for secret-key distillation for them. Hence,
if quantum channels $\{{\cal N}^e\}_{e\in E}$ in the network satisfy
\begin{align}
Q^\leftrightarrow (\{\bar{f}^e/\bar{f}^{V_A \leftrightarrow V_B}, {\cal N}^{e} \}_{e\in V_A \leftrightarrow V_B})\le \sum_{e\in V_A\leftrightarrow V_B} \frac{\bar{f}^e Q^\leftrightarrow ({\cal N}^{e})}{\bar{f}^{V_A \leftrightarrow V_B}}, \label{eq:con-1}
\end{align}
the aggregated quantum repeater protocol is optimal for the entanglement distribution.
Similarly, if quantum channels $\{{\cal N}^e\}_{e\in E}$ in the network have properties
\begin{align}
&K^\leftrightarrow (\{\bar{f}^e/\bar{f}^{V_A \leftrightarrow V_B}, {\cal N}^{e} \}_{e\in V_A \leftrightarrow V_B}) \le \sum_{e\in V_A\leftrightarrow V_B} \frac{\bar{f}^e K^\leftrightarrow ({\cal N}^{e})}{\bar{f}^{V_A \leftrightarrow V_B}}, \\
& K^\leftrightarrow ({\cal N}^{e})= Q^\leftrightarrow ({\cal N}^{e}),\label{eq:con-3}
\end{align}
the aggregated quantum repeater protocol is optimal even for the secret-key distillation. 
The conditions (\ref{eq:con-1})-(\ref{eq:con-3}) are satisfied by the quantum network composed of a wide range of stretchable quantum channels such as erasure channels, dephasing channels, bosonic quantum amplifier channels, and optical lossy channels in the asymptotic limit \cite{PLO15}, but may not be in general \cite{HHH99,HHHO05,HPHH08,SY08,BCHW15}.
However, it is a more important fact that
the optimality of the aggregated quantum repeater protocol is now related to fundamental questions on whether the given quantum channels satisfy fundamental relations (\ref{eq:con-1})-(\ref{eq:con-3}) or not.

We thank S.~B\"auml, H.-K.~Lo, A.~Mizutani, W.~J.~Munro, K.~Tamaki, and R. Van Meter for valuable comments.
This work was in part funded by ImPACT Program of Council for Science, Technology and Innovation (Cabinet Office, Government of Japan).

\end{document}